\documentclass{article}
\usepackage{spconf,amsmath,graphicx}
\usepackage{enumitem}
\setlist{nosep, leftmargin=14pt}

\usepackage{mwe} % to get dummy images
\usepackage[utf8]{inputenc}
\usepackage[T1]{fontenc}

\usepackage{import}
\usepackage{amsmath,graphicx}
\usepackage{times}
\usepackage[dvipsnames,table]{xcolor}
\usepackage{pgf}

\usepackage{tabularx,booktabs}
\usepackage{arydshln}
\usepackage{tabu}
\usepackage{adjustbox}
\usepackage{enumerate,multicol}
\usepackage{multirow}
\usepackage{multicol}
\usepackage{colortbl}
\definecolor{gradientstart}{rgb}{0.97, 0.98, 1} % Light blue shade start
\definecolor{gradientend}{rgb}{0.9, 0.95, 1}    % Slightly darker blue

\usepackage{float}
\usepackage{balance}
\usepackage{xspace}
\usepackage{amssymb}
\usepackage{bbding,enumitem}
\usepackage{booktabs}
\usepackage{array}
\usepackage{hyperref}

\usepackage{xcolor}
\definecolor{darkblue}{rgb}{0.0, 0.0, 0.5} % Dark blue color
\hypersetup{
    colorlinks=true,
    linkcolor=darkblue,
    filecolor=darkblue,
    urlcolor=darkblue,
    citecolor=darkblue
}
\usepackage{color, colortbl}
\hypersetup{colorlinks,linkcolor={red!50!black},citecolor={blue!50!black},urlcolor={black}}
\usepackage[table]{xcolor}
\usepackage{times}

\usepackage{soul}
\usepackage{url}
\usepackage{float}

\usepackage{amsmath}
\usepackage{nicefrac, xfrac}
\usepackage{booktabs}

\definecolor{gray1}{RGB}{210,210,210}
\definecolor{gray2}{RGB}{50,50,50}
\definecolor{gray3}{RGB}{150,150,150}
\definecolor{RYB1}{RGB}{218,232,252}
\definecolor{RYB4}{RGB}{108,142,191}
\definecolor{blue2}{RGB}{230,240,250}

\makeatletter
\newcommand\avsuminner[2]{%
  {\sbox0{$\m@th#1\sum$}%
   \vphantom{\usebox0}%
   \ooalign{%
     \hidewidth
     \smash{\vrule height\dimexpr\ht0+1pt\relax depth\dimexpr\dp0+1pt\relax}%
     \hidewidth\cr
     $\m@th#1\sum$\cr
   }%
  }%
}
\definecolor{shadecolor}{rgb}{0.9,0.9,0.9}

\usepackage{verbatim}
\usepackage{pifont}
\usepackage{tikz}

\newcommand{\DP}[2]{%
  \begin{tikzpicture}
    \fill[color=#2]   (0.0 , 0.0) rectangle (#1*7.8ex , 2ex );
  \end{tikzpicture}%
}

\usepackage{calc}
\usepackage{stackengine}
\newcommand\clapp[3][0pt]{\stackengine{0pt}{#3}{\kern#1#2}{O}{c}{F}{F}{L}}

\usepackage{pgfplots}
\usepackage{pgfplotstable}
\usetikzlibrary{pgfplots.groupplots}
\usepgfplotslibrary{colorbrewer}
\pgfplotsset{/pgfplots/error bars/error bar style={black,thick}}

\pgfplotsset{compat=1.11,
        /pgfplots/ybar legend/.style={
        /pgfplots/legend image code/.code={%
        \draw[##1,/tikz/.cd,bar width=3pt,yshift=-0.2em,bar shift=0pt]
                plot coordinates {(0cm,0.8em)};},
},}

\usepackage[linesnumbered,ruled]{algorithm2e}
\usepackage{todonotes}

\DeclareMathOperator*{\argmax}{argmax}
\usepackage{arydshln} % dashed table midrule (hdashline)

\usepackage[T1]{fontenc}

\graphicspath{{figures/}}
\usepackage{stackengine}

\usepackage{algorithm2e}
\usepackage{algpseudocode}
\usepackage{color}

\usepackage{pifont}
\usepackage{tikz}
\newcommand{\compareFOne}[2]{
    \pgfmathsetmacro{\diff}{#1-#2}
    \ifdim \diff pt > 0pt
        \(#1 \, \textcolor{green}{$\uparrow$ \diff}\)
    \else
        \(#1 \, \textcolor{red}{$\downarrow$ \diff}\)
    \fi
}

\title{Dual Invariance Self-Training for Reliable Semi-Supervised Surgical Phase Recognition}

\name{Sahar Nasirihaghighi$^{\star}$ \qquad Negin Ghamsarian$^{\dagger}$ \qquad Raphael Sznitman$^{\dagger}$ \qquad Klaus Schoeffmann$^{\star}$}

\address{$^{\star}$ Institute of Information Technology (ITEC), University of Klagenfurt, Austria \\
    $^{\dagger}$ Center for AI in Medicine, University of Bern, Switzerland}

\begin{document}
\maketitle

\begin{abstract}
Accurate surgical phase recognition is crucial for advancing computer-assisted interventions, yet the scarcity of labeled data hinders training reliable deep learning models. Semi-supervised learning (SSL), particularly with pseudo-labeling, shows promise over fully supervised methods but often lacks reliable pseudo-label assessment mechanisms. To address this gap,  we propose a novel SSL framework, \textbf{Dual Invariance Self-Training (DIST)}, that incorporates both \textbf{Temporal} and \textbf{Transformation Invariance} to enhance surgical phase recognition. Our two-step self-training process dynamically selects reliable pseudo-labels, ensuring robust pseudo-supervision. Our approach mitigates the risk of noisy pseudo-labels, steering decision boundaries toward true data distribution and improving generalization to unseen data. Evaluations on Cataract and Cholec80 datasets show our method outperforms state-of-the-art SSL approaches, consistently surpassing both supervised and SSL baselines across various network architectures. Code is available at {\small \textit{\href{https://github.com/Sahar-Nasiri/DIST}{\textcolor{darkblue}{https://github.com/Sahar-Nasiri/DIST}}}}.

\end{abstract}

\vspace{-1em}

\section{Introduction}
\label{sec:intro}
\vspace{-0.7em}
In the field of surgical video analysis, supervised deep learning has significantly advanced the accuracy of phase and action recognition \cite{nasirihaghighi2023action,ghamsarian2021relevance,nasirihaghighi2024event}. While effective, these methods rely on extensive labeled data, which can be costly and time-consuming, especially in complex fields like surgical video analysis, where expert annotations are required.

\noindent Semi-supervised learning (SSL) effectively enhances model performance when only limited labeled data is available, especially in areas like video action recognition. By using a small labeled dataset together with a larger pool of unlabeled data, SSL enables models to extract relevant patterns and generalize well, avoiding overfitting to the limited labeled set. Techniques such as consistency regularization~\cite{chen2021semi}, contrastive learning~\cite{chaitanya2020contrastive}, and self-training~\cite{yang2022st++} allow models to learn meaningful representations from unlabeled data, thereby improving model generalization and robustness.

\noindent While SSL has shown promise in action recognition, its application to surgical phase recognition remains underexplored. Recent advancements, such as transformer-based models optimized for low-data scenarios and innovative augmentations like Tube TokenMix for enhanced temporal learning~\cite{xing2023svformer}, highlight SSL's potential in video-based tasks. Furthermore, student-teacher frameworks have proven effective in semi-supervised action recognition~\cite{dave2023timebalance, xu2022cross}. Despite these advancements, we argue that more reliable pseudo-supervision can further enhance SSL. This motivates our study, where we present a novel SSL framework based on pseudo-supervision with temporal and transformation invariance. Our main contributions are as follows:

\begin{itemize}
    \item We introduce a novel SSL framework for phase recognition in surgical videos based on reliable pseudo-supervision. More specifically, we adopt a temporal and transformation invariant self-assessment mechanism for selecting reliable pseudo-labels, by which our framework can estimate and filter out the wrong pseudo-labels.
    
    \item We extensively evaluate our model on two surgical datasets using three advanced network architectures and compare it against state-of-the-art SSL methods for action recognition.
 
    \item The experimental results show promising performance of the proposed method under different portions of available labeled data with different network architectures against all baseline methods.
\end{itemize}

\vspace{-1em}

\section{Proposed Approach}
\label{sec:approach}
\vspace{-0.7em}
\begin{figure}[!t]
    \centering
    % \captionsetup{font=small}
    \includegraphics[width=0.44\textwidth]{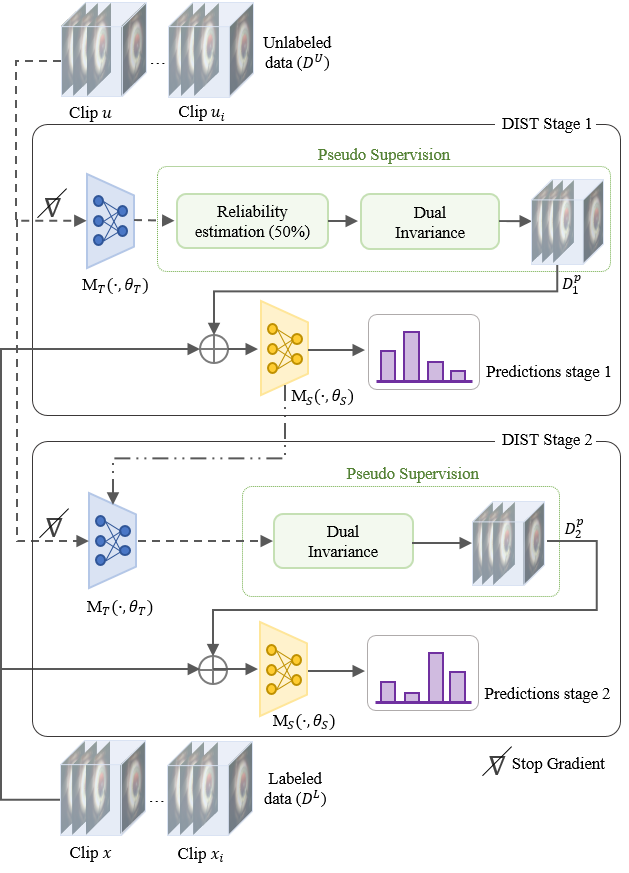}
    \caption{The framework of the proposed model}
    \label{fig:Overview}
\end{figure}
\textbf{Overview.} Our semi-supervised self-training framework, Figure \ref{fig:Overview}, operates in two stages: first, a teacher model is trained on labeled data and generates pseudo-labels for unlabeled data, selecting the top $50\%$ pseudo-label based on our proposed reliability estimation method. We then apply our dual invariance pseudo-label consistency measurement strategy to filter out further unreliable predictions, retaining only the most robust pseudo-labels. The retained pseudo labels, combined with labeled data, train a student model. This student then becomes the new teacher in the second stage, generating more refined pseudo labels. In the second stage, all pseudo labels undergo our proposed consistency measurement method, and the retained pseudo labels are mixed with labeled data to train a student model. Algorithm \ref{algorithm} provides a detailed step-by-step overview of this process, which is further explained in the following subsections.

\vspace{-0.5em}
\textbf{Problem Definition.} Given a labeled dataset $\mathcal{D^L}$ with $\mathcal{N^L}$ video clips $(x_i, y_i)$, where $x_i$ is a video clip and $y_i$ its correcponding phase label, and an unlabeled dataset $\mathcal{D^U}$ with $\mathcal{N^U}$ clips $(u_i)$ (where $\mathcal{N^U} >> \mathcal{N^L}$), the aim is to train a model $\mathcal{M_S}(\cdot, \theta_s)$ that utilizes both $\mathcal{D^L}$ and $\mathcal{D^U}$ for better performance on unseen surgical videos.

\vspace{-0.5em}
\textbf{Dual Invariance Filtering.} To estimate the unreliability of pseudo labels, we propose to exploit a prior knowledge that the network is expected to learn during training. More concretely, a trained neural network for phase recognition is expected to provide consistent predictions for different samples of the same sequence, irrespective of temporal frame sampling or frame-wise transformations. Indeed, inconsistent predictions of a network for different versions of identical inputs are a clue of unreliability of predictions. Hence, we feed two versions of each unlabeled video to the teacher model: (1) the uniformly sampled sequence ($u_i^f$) and (2) a randomly sampled and strongly augmented counterpart ($u_i^g = h(u_i^g, \phi_h)$), representing the temporal and transformation invariance conditions. We then filter out sequences corresponding to inconsistent predictions as follows:

\begin{algorithm}[t!]
\footnotesize
    \SetKwInOut{Input}{Input}
    \SetKwInOut{Output}{Output}
    
    \Input{Labeled dataset $\mathcal{D^L}$, Unlabeled dataset $\mathcal{D^U}$\\
    Pseudo-label sets: $\mathcal{D}{_1}{^p}, \mathcal{D}{_2}{^p}, \mathcal{D}^{mix}  = \varnothing$\\
    Uniform and random frame sampling\\function $f(\cdot, \phi_f)$, $g(\cdot, \phi_g)$\\
    Data augmentation $h(\cdot, \phi_h)$\\
    Teacher model $\mathcal{M}_{T}(\cdot, \theta_T)$, Student model $\mathcal{M}_{S}(\cdot, \theta_S)$}
    
    \Output{Trained student model $\mathcal{M}_{S}(\cdot, \theta_S)$}
    
    \For{video $u_i$ in $\mathcal{D^U}$}{
        Uniform frame sampling $u_i^f = f(u_i, \phi_f)$\\
        Generate pseudo-label $\hat{z}_i$ using $\mathcal{M}_{T}$ and estimate its reliability. 
        \Comment{Refer to Eq. \eqref{eq:pseudo} and \eqref{eq:consistency}} \\
        $\mathcal{D}{_1}{^{p}} = \text{Score} (u_i^f, \hat{z}_i) > \text{Mdn} (\mathcal{D}{_1}{^p})$
        \Comment{Top 50\% selection}\\
        Apply dual invariance filtering on $\mathcal{D}{_1}{^{p}}$ \Comment{Refer to Eq. \eqref{eq:DI}}
    }
    
    $\mathcal{D}^{mix} = \mathcal{D^L} \cup \mathcal{D}{_1}{^p}$\\
    Train $\mathcal{M}_{S}$ on $\mathcal{D}^{mix}$ using $g(\cdot, \phi_g)$ and $h(\cdot, \phi_h)$\\
    $\mathcal{M}_{T} \leftarrow \mathcal{M}_{S}$ \\
    \For{video $u_i$ in $\mathcal{D^U}$}{
        Uniform frame sampling $u_i^f = f(u_i, \phi_f)$\\
        Generate pseudo-label $\hat{z}_i$ using $\mathcal{M}_{T}$ \Comment{Refer to Eq. \eqref{eq:pseudo}}\\
        %(see Eq. \ref{eq:pseudo})
        $\mathcal{D}{_2}{^{p}}$ = All generated pseudo-labels\\
        Apply dual invariance filtering on $\mathcal{D}{_2}{^{p}}$ 
        \Comment{Refer to Eq. \eqref{eq:DI}}
    }
    
    $\mathcal{D}^{mix} = \mathcal{D^L} \cup \mathcal{D}{_2}{^p}$\\
    Train $\mathcal{M}_{S}$ on $\mathcal{D}^{mix}$ using $g(\cdot, \phi_g)$ and $h(\cdot, \phi_h)$
    
    \caption{DIST}
    \label{algorithm}
\end{algorithm}
\vspace{-1em}

\vspace{-1em}

\begin{equation}
\mathcal{D^U} \leftarrow 
\begin{cases} 
(u_i, \mathcal{M}_{T}(u_i^f,\theta_T)) & \text{if } \mathcal{M}_{T}(u_i^f,\theta_T) = \mathcal{M}_{T}(u_i^g,\theta_T) \\
\text{ignore} & \text{else}
\end{cases}
\label{eq:DI}
\end{equation}

\textbf{DIST Stage 1.} In the initial step, we train a teacher model $\mathcal{M}_{T}(\cdot, \theta_T)$ using only the labeled dataset $\mathcal{D^L}$ for $n$ epochs. Inspired by ST++~\cite{yang2022st++}, we evaluate the reliability of pseudo-labels based on the consistency of predictions over time. To do this, we save three checkpoints $\big{(}T(\frac{n}{3}), T(\frac{2n}{3}), T(n)\big{)}$ at intervals during training, which are then used to generate three sets of pseudo-labels for the unlabeled dataset. This approach allows us to assess the consistency of the pseudo-labels for each video clip in the unlabeled set. To determine reliability, we calculate a score across three saved checkpoints to measure pseudo-label consistency over time as follows:
\vspace{-0.1em}
\begin{align}
z_i^n = & \sigma(\mathcal{M}_{T}(u_i^f, \theta_T(n))) \notag \\ 
\hat{z}_i^n = &\argmax_C(z_i^n)
\label{eq:pseudo}
\end{align}
\vspace{-0.5em}
\begin{equation}
\mathcal{R}_{u_i} = \Big(\frac{z_i^{\frac{n}{3}} \times \hat{z}_i^{n}}{z_i^{\frac{n}{3}} + \hat{z}_i^{n}} + 2\times\frac{z_i^{\frac{2n}{3}} \times \hat{z}_i^{n}}{z_i^{\frac{2n}{3}} + \hat{z}_i^{n}}\Big)/3
\label{eq:consistency}
\end{equation}

%%%%%%%%%%%%%%%%%%%%%%%%%%%%%%%%%%%%%%%%%%%%%%%%%%%%%
% Cataract dataset- VGG_Transformer- VGG_LSTM - ResNet3D Accuracy/ F1-score
\begin{table*}[!ht]
\renewcommand{\arraystretch}{0.5}
\centering
% \captionsetup{font=small}
\small
\caption{Comparative analysis of various models across three networks, evaluated using Accuracy and F1-Score metrics on the Cataract-1k dataset. The best and second-best values are highlighted in green and blue, respectively. }
\label{tab:acc_cat}
\resizebox{1\textwidth}{!}{%

\begin{tabular}{>{\centering\arraybackslash}m{0.85cm}*{1}{>{\raggedright\arraybackslash}m{1.45cm}}*{5}{>{\centering\arraybackslash}m{3.4cm}}>{\centering\arraybackslash}m{1.3cm}}

\toprule
& & 1/2 (4263) & 1/4 (2111) & 1/8 (1323) & 1/16 (710) & 1/32 (481) & Average\\ 
\cmidrule(lr){3-3}\cmidrule(lr){4-4}\cmidrule(lr){5-5}\cmidrule(lr){6-6}\cmidrule(lr){7-7}\cmidrule(lr){8-8}
Network & Model & Acc|F1 & Acc|F1 & Acc|F1 & Acc|F1 & Acc|F1 & Accuracy\\\midrule
\multirow{15}{*}{\rotatebox{90}{Transformer}}&Supervised&94.45 || 94.44&90.94 || 90.88&87.67 || 87.46&84.70 || 84.35&77.43 || 76.12&N/A\\
&FixMatch&96.45 || 96.42&95.09 || 95.04&93.01 || 92.42&90.97 || 90.29&83.67 || 81.45&4.80\\
&MeanTeacher& 96.01 || 96.03&93.96 || 93.93&84.93 || 83.38&77.80 || 75.50&54.21 || 51.49&-5.66\\
&CMPL&96.89 || 96.85&96.73 || 96.68&91.63 || 90.77&83.26 || 79.87&72.47 || 60.98&1.16\\
&SVFormer&\textbf{\textcolor{Green}{97.98}} || \textbf{\textcolor{Green}{97.93}}&96.74 || 96.33&92.20 || 91.29&84.04 || 80.38&73.55 || 67.14&1.88\\
&\cellcolor{gradientstart} ST\_stage1&\cellcolor{gradientstart} \textbf{\textcolor{Cerulean}{97.93}} || \cellcolor{gradientstart} 97.90&\cellcolor{gradientstart} 96.52 || \cellcolor{gradientstart} 96.48&\cellcolor{gradientstart} \textbf{\textcolor{Cerulean}{94.47}} || \cellcolor{gradientstart} \textbf{\textcolor{Cerulean}{94.18}}& \cellcolor{gradientstart}91.76 || \cellcolor{gradientstart}90.68& \cellcolor{gradientstart}\textbf{\textcolor{Cerulean}{86.14}} || \cellcolor{gradientstart}\textbf{\textcolor{Cerulean}{84.03}}& \cellcolor{gradientstart}6.32\\

&\cellcolor{gradientstart}ST\_stage2&
\cellcolor{gradientstart}97.90 || \cellcolor{gradientstart}\textbf{\textcolor{Cerulean}{97.89}}&
\cellcolor{gradientstart}\textbf{\textcolor{Cerulean}{97.06}} || \cellcolor{gradientstart}\textbf{\textcolor{Cerulean}{97.04}}&
\cellcolor{gradientstart}94.27 || \cellcolor{gradientstart}93.95&
\cellcolor{gradientstart}\textbf{\textcolor{Cerulean}{92.11}} || \cellcolor{gradientstart}\textbf{\textcolor{Cerulean}{90.82} }&
\cellcolor{gradientstart}\textbf{\textcolor{Green}{86.60}} || \cellcolor{gradientstart}\textbf{\textcolor{Green}{84.22}}&
\cellcolor{gradientstart}\textbf{\textcolor{Cerulean}{6.55}}\\

&\cellcolor{gradientend}DIST\_stage1& \cellcolor{gradientend}97.70 || \cellcolor{gradientend}97.68& \cellcolor{gradientend}96.69 || \cellcolor{gradientend}96.65& \cellcolor{gradientend}94.17 || \cellcolor{gradientend}93.90& \cellcolor{gradientend}91.22 || \cellcolor{gradientend}90.16& \cellcolor{gradientend}85.93 || \cellcolor{gradientend}84.02& \cellcolor{gradientend}6.10\\

&\cellcolor{gradientend}DIST\_stage2&
\cellcolor{gradientend}97.74 || \cellcolor{gradientend}97.73 \makebox[0pt][l]{\raisebox{0.1ex}{\textcolor{gray}{(3.29 $\uparrow$)}}}&
\cellcolor{gradientend}\textbf{\textcolor{Green}{97.47}} || \cellcolor{gradientend}\textbf{\textcolor{Green}{97.46}} \makebox[0pt][l]{\raisebox{0.1ex}{\textcolor{gray}{(6.58$\uparrow$)}}}&
\textbf{\textcolor{Green}{94.74}} || \cellcolor{gradientend}\textbf{\textcolor{Green}{94.48}} \makebox[0pt][l]{\raisebox{0.1ex}{\textcolor{gray}{(7.02$\uparrow$)}}}&
\cellcolor{gradientend}\textbf{\textcolor{Green}{92.78}} || \cellcolor{gradientend}\textbf{\textcolor{Green}{91.60}} \makebox[0pt][l]{\raisebox{0.1ex}{\textcolor{gray}{(7.25$\uparrow$)}}}&
\cellcolor{gradientend}85.97 || \cellcolor{gradientend}83.62 \makebox[0pt][l]{\raisebox{0.1ex}{\textcolor{gray}{(7.5$\uparrow$)}}}&
\cellcolor{gradientend}\textbf{\textcolor{Green}{6.70}}\\
\hdashline

\multirow{13}{*}{\rotatebox{90}{VGG-LSTM}}&Supervised&96.51 || 96.50&94.06 || 94.03&92.30 || 92.14&87.09 || 86.97&81.28 || 80.37&N/A\\
&FixMatch&96.89 || 96.86&96.37 || 96.34&\textbf{\textcolor{Cerulean}{94.61}} || \textbf{\textcolor{Cerulean}{94.45}}&90.62 || 90.01&86.05 || 84.53&2.66\\

&MeanTeacher&92.72 || 92.66&89.65 || 89.50&84.86 || 83.67&78.60 || 76.88&73.40 || 70.66&-6.40\\

&CMPL&97.26 || 97.22&96.46 || 96.61&93.67 || 92.19&85.53 || 84.20&75.15 || 67.36&-0.63\\

&\cellcolor{gradientstart}ST\_stage1&\cellcolor{gradientstart}\textbf{\textcolor{Cerulean}{97.94}} || \cellcolor{gradientstart}97.92&96.56 || \cellcolor{gradientstart}96.54&94.42 || \cellcolor{gradientstart}94.16&91.08 || \cellcolor{gradientstart}90.66&\textbf{\textcolor{Cerulean}{86.12}} || \cellcolor{gradientstart}\textbf{\textcolor{Cerulean}{84.56}}& \cellcolor{gradientstart}2.97\\

&\cellcolor{gradientstart}ST\_stage2&
\cellcolor{gradientstart}97.89 || \cellcolor{gradientstart}\textbf{\textcolor{Cerulean}{97.88}}&
\cellcolor{gradientstart}\textbf{\textcolor{Cerulean}{97.25}} || \cellcolor{gradientstart}\textbf{\textcolor{Cerulean}{97.23}}&
\cellcolor{gradientstart}94.28 || \cellcolor{gradientstart}93.90&
\cellcolor{gradientstart}\textbf{\textcolor{Green}{91.40}} || \cellcolor{gradientstart}\textbf{\textcolor{Cerulean}{90.79}}&
\cellcolor{gradientstart}86.03 || \cellcolor{gradientstart}84.22&
\cellcolor{gradientstart}\textbf{\textcolor{Cerulean}{3.12}}\\

&\cellcolor{gradientend}DIST\_stage1&\cellcolor{gradientend}97.76 || \cellcolor{gradientend}97.74& \cellcolor{gradientend}96.64 || \cellcolor{gradientend}96.62&\textbf{\textcolor{Green}{95.05}} || \cellcolor{gradientend}\textbf{\textcolor{Green}{94.84}}& \cellcolor{gradientend}91.09 || \cellcolor{gradientend}90.73& \cellcolor{gradientend}85.92 || \cellcolor{gradientend}84.40& \cellcolor{gradientend}3.04\\

&\cellcolor{gradientend}DIST\_stage2&
\cellcolor{gradientend}\textbf{\textcolor{Green}{98.09}} || \cellcolor{gradientend}\textbf{\textcolor{Green}{98.08}} \makebox[0pt][l]{\raisebox{0.1ex}{\textcolor{gray}{(1.58$\uparrow$)}}} &
\cellcolor{gradientend}\textbf{\textcolor{Green}{97.27}} || \cellcolor{gradientend}\textbf{\textcolor{Green}{97.24}} \makebox[0pt][l]{\raisebox{0.1ex}{\textcolor{gray}{(3.21$\uparrow$)}}}&
\cellcolor{gradientend}94.47 || \cellcolor{gradientend}94.15 \makebox[0pt][l]{\raisebox{0.1ex}{\hspace{0.2em}\textcolor{gray}{(2.01$\uparrow$)}}}&
\cellcolor{gradientend}\textbf{\textcolor{Cerulean}{91.31}} || \cellcolor{gradientend}\textbf{\textcolor{Green}{90.84}} \makebox[0pt][l]{\raisebox{0.1ex}{\textcolor{gray}{(3.87$\uparrow$)}}}&
\cellcolor{gradientend}\textbf{\textcolor{Green}{86.65}} || \cellcolor{gradientend}\textbf{\textcolor{Green}{84.91}} \makebox[0pt][l]{\raisebox{0.1ex}{\textcolor{gray}{(4.54$\uparrow$)}}}&
\cellcolor{gradientend}\textbf{\textcolor{Green}{3.31}}\\
\hdashline

\multirow{15}{*}{\rotatebox{90}{ResNet3D}}&Supervised&96.55 || 96.55&93.83 || 93.82&87.19 || 86.95&83.37 || 82.90&72.21 || 70.53&N/A\\

&FixMatch&96.84 || 97.14&\textbf{\textcolor{Green}{96.51}} || 95.18&89.60 || 86.42&85.95 || \textbf{\textcolor{Cerulean}{85.64}}&78.99 || 77.45&2.95\\

&MeanTeacher&87.26 || 88.09&85.28 || 86.67&80.68 || 83.34&77.49 || 80.00&68.87 || 69.55&-6.71\\

&CMPL&97.57 || 97.47&96.12 || 93.90&88.92 || 88.18&83.81 || 83.39&72.87 || 71.17&1.23\\

&TG&95.19 || 95.24&90.34 || 87.09&81.45 || 79.56&72.67 || 73.81&58.88 || 50.64&-6.92\\

&\cellcolor{gradientstart}ST\_stage1& \cellcolor{gradientstart}97.78 || \cellcolor{gradientstart}97.79& \cellcolor{gradientstart}95.64 || \cellcolor{gradientstart}95.61& \cellcolor{gradientstart}90.41 || \cellcolor{gradientstart}90.28& \cellcolor{gradientstart}84.70 || \cellcolor{gradientstart}83.84& \cellcolor{gradientstart}78.20 || \cellcolor{gradientstart}76.77& \cellcolor{gradientstart}2.72\\
&\cellcolor{gradientstart}ST\_stage2&
\cellcolor{gradientstart}\textbf{\textcolor{Cerulean}{98.33}} || \cellcolor{gradientstart}\textbf{\textcolor{Cerulean}{98.33}}&
\cellcolor{gradientstart}96.00 || \cellcolor{gradientstart}\textbf{\textcolor{Cerulean}{95.96}}&
\cellcolor{gradientstart}\textbf{\textcolor{Cerulean}{91.61}} || \cellcolor{gradientstart}91.33&
\cellcolor{gradientstart}80.56 || \cellcolor{gradientstart}76.04&
\cellcolor{gradientstart}74.69 || \cellcolor{gradientstart}69.61&\cellcolor{gradientstart}1.61\\
&\cellcolor{gradientend}DIST\_stage1&\cellcolor{gradientend}97.54 || \cellcolor{gradientend}97.54& \cellcolor{gradientend}95.74 || \cellcolor{gradientend}95.73& \cellcolor{gradientend}91.48 || \cellcolor{gradientend}\textbf{\textcolor{Cerulean}{91.35}}&\cellcolor{gradientend}\textbf{\textcolor{Cerulean}{86.52}} || \cellcolor{gradientend}\textbf{\textcolor{Cerulean}{85.64}}& \cellcolor{gradientend}\textbf{\textcolor{Cerulean}{80.33}} || \cellcolor{gradientend}\textbf{\textcolor{Cerulean}{78.87}}& \cellcolor{gradientend}\textbf{\textcolor{Cerulean}{3.69}}\\
&\cellcolor{gradientend}DIST\_stage2&
\cellcolor{gradientend}\textbf{\textcolor{Green}{98.79}} || \cellcolor{gradientend}\textbf{\textcolor{Green}{98.79}} \makebox[0pt][l]{\raisebox{0.1ex}{\textcolor{gray}{(2.24$\uparrow$)}}}&
\cellcolor{gradientend}\textbf{\textcolor{Cerulean}{96.39}} || \cellcolor{gradientend}\textbf{\textcolor{Green}{96.37}} \makebox[0pt][l]{\raisebox{0.1ex}{\textcolor{gray}{(2.55$\uparrow$)}}}&
\cellcolor{gradientend}\textbf{\textcolor{Green}{92.29}} || \cellcolor{gradientend}\textbf{\textcolor{Green}{92.18}} \makebox[0pt][l]{\raisebox{0.1ex}{\textcolor{gray}{(5.23$\uparrow$)}}}&
\cellcolor{gradientend}\textbf{\textcolor{Green}{87.24}} || \cellcolor{gradientend}\textbf{\textcolor{Green}{86.04}} \makebox[0pt][l]{\raisebox{0.1ex}{\textcolor{gray}{(3.14$\uparrow$)}}}& \cellcolor{gradientend}\textbf{\textcolor{Green}{82.33}} || \cellcolor{gradientend}\textbf{\textcolor{Green}{81.12}} \makebox[0pt][l]{\raisebox{0.1ex}{\textcolor{gray}{(10.59$\uparrow$)}}}&
\cellcolor{gradientend}\textbf{\textcolor{Green}{4.78}}\\ \midrule

\end{tabular}
}
\end{table*}

%\vspace{-1em}
\noindent Where $\sigma$ is the Softmax operation. The final network predictions are treated as hard labels, while earlier predictions are weighted by their temporal distance, reflecting the model’s learning evolution. We retain the top 50\% of pseudo-labels with the highest reliability scores, which then undergo dual invariance filtering (as described in Eq. \ref{eq:DI}). This ensures that only pseudo-labels invariant to temporal sampling and transformation are retained. The resulting set of reliable pseudo-labels $\mathcal{D}{_1}{^p}$, combined with the original labeled dataset $\mathcal{D^L}$, is used to train the student model $\mathcal{M_S}(\cdot, \theta_s)$, which is pre-trained on ImageNet~\cite{ImageNet} to promote better adaptation and prevent early convergence.

\textbf{DIST Stage 2.} In this stage, the student model from stage 1 becomes the new teacher model to generate refined pseudo-labels for all unlabeled samples $\mathcal{D^U}$. Using the improved predictions, we only reapply the dual invariant filtering to select the most reliable pseudo-labels $\mathcal{D}{_2}{^p}$, which are then combined with the original labeled data, $\mathcal{D^L}$, to train a student model.

\vspace{-1.5em}

\section{Experimental Settings}
\label{sec:settings}
\vspace{-0.85em}

\textbf{Datasets.} 
We evaluate our approach on subsets of the Cataract-1k dataset~\cite{ghamsarian2024cataract} and Cholec80~\cite{twinanda2016endonet}, using 42 annotated videos each. For Cataract-1k, we focus on three phases (\textit{irrigation-aspiration}, \textit{lens implantation}, and \textit{phacoemulsification}) and group other phases as ``rest''. In Cholec80, we target \textit{CalotTriangleDissection}, \textit{Preparation}, and \textit{GallbladderPackaging}. We use 36 videos for training and 4 for testing, dividing each phase into three-second clips, resulting in 8,857 training clips for Cataract-1k and 7,171 for Cholec80. To assess performance across different labeled-to-unlabeled ratios, we split the training set by randomly sub-sampling 1/32, 1/16, 1/8, 1/4, and 1/2 of the training videos as labeled, with the remainder serving as the unlabeled set.

\textbf{Baseline models. }We compare our proposed semi-supervised framework with state-of-the-art baselines, including FixMatch~\cite{sohn2020fixmatch}, Mean Teacher~\cite{tarvainen2017mean}, Cross-Model Pseudo Labeling (CMPL)~\cite{xu2022cross}, SVFormer~\cite{xing2023svformer}, and Temporal Gradient (TG)~\cite{xiao2022learning}.

\textbf{Networks and Training Settings.} We extensively evaluate the proposed and baseline methods using three network architectures: a hybrid transformer \cite{djenouri2023hybrid} combining VGG16 with a transformer \cite{nasirihaghighi2024event}, a VGG-LSTM model \cite{nasirihaghighi2023action, ghamsarian2020enabling} (both backbones pre-trained on ImageNet), and a ResNet3D-18 model \cite{hara2017learning} pre-trained on natural videos. Training is performed with a batch size of 16 for 40 epochs, starting with a learning rate of 0.005 that decays by $\gamma=0.9$ every other epoch, using the SGD optimizer with 0.9 momentum. The input size is set to $256\times 256$ pixels, and data augmentation includes random rotations (up to 15 degrees), color jittering (brightness=0.3, contrast=0.3, saturation=0.5), and Gaussian blurring with a kernel size of five and sigma ranging from 0.1 to 2.0. The models are trained using cross-entropy loss, and we evaluate them using four-fold cross-validation, averaging results across all folds. The proposed model adds no extra testing time compared to supervised models.

% Notably, the proposed model achieves its performance without incurring additional testing time compared to supervised models.

\vspace{-1.5em}

\section{Experimental Results}
\label{sec:results}
\vspace{-0.85em}

Table \ref{tab:acc_cat} shows the performance comparison of three different networks on the Cataract-1k dataset across five data splits, using accuracy and F1-score metrics. Our proposed model (DIST\_stage1 and stage2) consistently outperforms both supervised and state-of-the-art semi-supervised methods across all networks, particularly excelling in data-limited scenarios. For example, in the 1/4 split, it surpasses the supervised model trained on twice as much labeled data (1/2 split) in both hybrid transformer and VGG\_LSTM networks. Remarkably, in the hybrid transformer, DIST\_stage2 achieves an impressive 92.78\% accuracy with only 710 labeled videos in the 1/16 split, surpassing the supervised model’s accuracy in the 1/4 split (90.94\%) with four times more labeled data (2,111 labeled videos). Even in the smallest split (1/32), where some frameworks fail to improve the performance, our model maintains robust performance across all network architectures. The reason why some models are susceptible to very-low data settings, is the lack of a pseudo-label reliability estimation strategy leading to error propagation by training on wrong pseudo labels. The last column in this table shows the average accuracy improvement of each model compared to the supervised baseline, with DIST\_stage2 consistently leading, especially in the hybrid transformer model with a 6.7\% average improvement. 

Models like TG, which is optimized for rapid action transitions, perform poorly on surgical videos due to dynamic backgrounds and subtle actions. These background changes mislead the model, causing it to focus on irrelevant motion rather than key surgical transitions, thus reducing its accuracy in phase recognition.

%%%%%%%%%%%%%%%%%%%%%%%%%%%%%%%%%%%%%%%%%%%%%%%%%
% Cholec80 dataset- VGG_Transformer- Accuracy/ F1-score 
\begin{table}[t]
\centering
% \captionsetup{font=small}

\caption{Comparative analysis of models on a hybrid transformer network, evaluated by Accuracy and F1-Score for the Cholec80 dataset.}
\label{tab:acc_Tran_Cholec}
\resizebox{\columnwidth}{!}{ % 
\begin{tabular}{l>{\raggedright\arraybackslash}p{3cm}>{\raggedright\arraybackslash}p{3cm}>{\raggedright\arraybackslash}p{3cm}>{\centering\arraybackslash}p{1.3cm}} 
\toprule
 & 1/8 (1306) & 1/16 (777) & 1/32 (517) & Average\\ \cmidrule(lr){2-2}\cmidrule(lr){3-3}\cmidrule(lr){4-4}\cmidrule(lr){5-5}
Model & Acc | F1 & Acc | F1 & Acc | F1 & Accuracy\\  \midrule
Supervised & 79.43 || 79.36 & 75.00 || 75.05 & 67.50 || 67.32 & N/A\\

FixMatch & \textbf{\textcolor{Green}{81.86}} || \textbf{\textcolor{Green}{81.75}} & 78.94 || 79.17 & 70.65 || 69.06 & 3.17\\

MeanTeacher & 79.11 || 77.86 & 70.86 || 67.53 & 54.84 || 52.07 & -5.71\\

CMPL & 80.55 || 75.83 & 74.45 || 56.69 & 63.78 || 57.88 & -1.05\\

SVFormer & 80.19 || 78.88 & 75.84 || 70.21 & 64.26 || 60.51 & -0.55\\

\rowcolor{gradientstart}ST\_stage1 & 80.78 || \textbf{\textcolor{Cerulean}{80.57}} & 79.15 || 79.03 & 73.87 || 72.71 & 3.96\\

\rowcolor{gradientstart}ST\_stage2 & 80.23 || 79.95 & 79.95 || \textbf{\textcolor{Cerulean}{79.86}} & \textbf{\textcolor{Green}{75.80}} || \textbf{\textcolor{Cerulean}{73.94}} & \textbf{\textcolor{Cerulean}{4.68}}\\

\rowcolor{gradientend}DIST\_stage1 & 79.82 || 79.66 & \textbf{\textcolor{Cerulean}{79.52}} || 79.43 & 74.63 || 73.44 & 4.01\\

\rowcolor{gradientend}DIST\_stage2 &
\textbf{\textcolor{Cerulean}{80.96}} || 80.53 \makebox[0pt][l]{\raisebox{0.1ex}{\textcolor{gray}{(1.17$\uparrow$)}}} &
\textbf{\textcolor{Green}{79.95}} || \textbf{\textcolor{Green}{79.97}} \makebox[0pt][l]{\raisebox{0.1ex}{\textcolor{gray}{(4.92$\uparrow$)}}} &
\textbf{\textcolor{Cerulean}{75.46}} || \textbf{\textcolor{Green}{74.31}}\makebox[0pt][l]{\raisebox{0.1ex}{\hspace{0.1em}\textcolor{gray}{(6.99$\uparrow$)}}} &
\textbf{\textcolor{Green}{4.82}}\\

\midrule
\end{tabular}
}
\end{table}

Table \ref{tab:acc_Tran_Cholec} compares the performance of the VGG-Transformer network on the Cholec80 dataset, selected for its strong results on the Cataract-1k dataset. This evaluation emphasizes smaller splits with minimal labeled data, presenting a more challenging scenario. Our proposed model consistently outperforms the supervised model, even when the latter uses twice the labeled data. While CMPL and SVFormer perform well on larger splits (1/2 and 1/4), their effectiveness decreases significantly in smaller splits (1/16 and 1/32). In contrast, our model remains robust across all splits, highlighting its effectiveness in challenging scenarios.

\begin{figure}[t]
    \centering
    % \captionsetup{font=small}
    \includegraphics[width=0.5\textwidth]{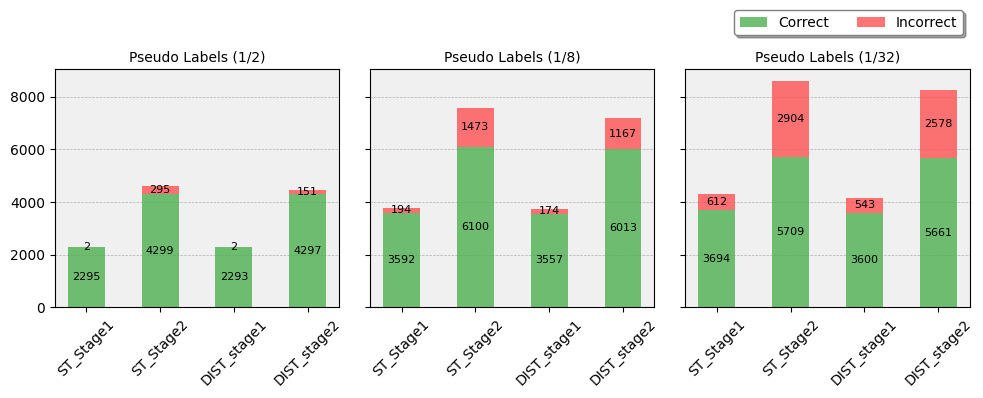}
    \caption{Comparison of the number of pseudo labels in the proposed model.}
    \label{fig:comparison}
\end{figure}

\textbf{Ablation Study. }We conducted an ablation study to evaluate the impact of pseudo-label selection within our DIST model. In the original DIST model, temporal and transformation invariance filtering (Section 3) is used in both stages to select the most reliable pseudo-labels. For the ablation, we removed this filtering, selecting $50\%$ of the most reliable pseudo-labels in the first stage (ST\_stage1) and retaining all pseudo-labels in the second stage (ST\_stage2), to examine its effect on pseudo-label reliability. 
As shown in the results tables, dual invariance filtering consistently improves performance across all networks by enhancing pseudo-label reliability, despite reducing the overall number of pseudo-labels. Figure \ref{fig:comparison} illustrates the correct and incorrect pseudo-labels generated in each stage for both the ablated (ST) and DIST models across three data splits. In the 1/2 split, ST produces nearly the same number of pseudo-labels as DIST, but stage 2 shows fewer incorrect labels. This improvement is even more notable in the 1/8 and 1/32 splits; while correct pseudo-labels slightly decrease, the substantial reduction in incorrect labels enhances pseudo-label reliability, ultimately improving the DIST model’s performance in surgical phase recognition.

%%%%%%%%%%%%%%%%%%%%%%%%%%%%%%%%%%%%%%%%%%%%%%%%%%%%%%%%%%
% Phase prediction performance of various models using hybrid transformer network.
\begin{table}[t]
\centering
% \captionsetup{font=small}
\caption{Phase prediction performance of various models using the hybrid transformer network on the Cataract-1k dataset.}
\label{tab:phase-annotations}
\resizebox{0.5\textwidth}{!}{%

\begin{tabular}{lll}
\specialrule{.12em}{.05em}{.05em}
Split & Model & Phases\\\midrule

\multicolumn{2}{c}{Ground-Truth} & \DP{2.00}{Gray}\DP{0.18}{Gray}\DP{3.04}{RoyalBlue}\DP{0.25}{Gray}\DP{1.17}{Dandelion}\DP{0.09}{Gray}\DP{0.48}{Gray}\DP{0.09}{Gray}\DP{0.27}{Purple}\DP{0.11}{Gray}\DP{0.23}{Gray}\DP{0.06}{Gray}\DP{0.50}{Dandelion}\DP{0.01}{Gray}\DP{0.11}{Dandelion}\DP{0.07}{Gray}\DP{0.40}{Gray}\DP{0.05}{Gray}\DP{0.25}{Dandelion}\DP{0.05}{Gray}\DP{0.59}{Gray}\\\hdashline
\multirow{6}{*}{1/4}&Supervised& \DP{2.04}{Gray}\DP{0.00}{Gray}\DP{0.08}{RoyalBlue}\DP{0.00}{Gray}\DP{0.08}{Purple}\DP{0.00}{Gray}\DP{0.23}{Gray}\DP{0.00}{Gray}\DP{0.08}{Dandelion}\DP{0.00}{Gray}\DP{0.08}{Gray}\DP{0.00}{Gray}\DP{0.15}{Dandelion}\DP{0.00}{Gray}\DP{2.49}{RoyalBlue}\DP{0.00}{Gray}\DP{0.30}{Gray}\DP{0.00}{Gray}\DP{1.13}{Dandelion}\DP{0.00}{Gray}\DP{0.75}{Gray}\DP{0.00}{Gray}\DP{0.15}{Purple}\DP{0.00}{Gray}\DP{0.45}{Gray}\DP{0.00}{Gray}\DP{0.60}{Dandelion}\DP{0.00}{Gray}\DP{0.53}{Gray}\DP{0.00}{Gray}\DP{0.30}{Dandelion}\DP{0.00}{Gray}\DP{0.53}{Gray}\DP{0.00}{Gray}\DP{0.04}{Purple}\\
&FixMatch&\DP{1.96}{Gray}\DP{0.00}{Gray}\DP{0.98}{Dandelion}\DP{0.00}{Gray}\DP{1.36}{RoyalBlue}\DP{0.00}{Gray}\DP{0.08}{Dandelion}\DP{0.00}{Gray}\DP{0.83}{RoyalBlue}\DP{0.00}{Gray}\DP{0.08}{Dandelion}\DP{0.00}{Gray}\DP{0.23}{Gray}\DP{0.00}{Gray}\DP{1.13}{Dandelion}\DP{0.00}{Gray}\DP{0.75}{Gray}\DP{0.00}{Gray}\DP{0.23}{Purple}\DP{0.00}{Gray}\DP{0.38}{Gray}\DP{0.00}{Gray}\DP{0.60}{Dandelion}\DP{0.00}{Gray}\DP{0.53}{Gray}\DP{0.00}{Gray}\DP{0.30}{Dandelion}\DP{0.00}{Gray}\DP{0.57}{Gray}\\
&MeanTeacher&\DP{1.96}{Gray}\DP{0.00}{Gray}\DP{0.15}{Dandelion}\DP{0.00}{Gray}\DP{3.09}{RoyalBlue}\DP{0.00}{Gray}\DP{0.30}{Gray}\DP{0.00}{Gray}\DP{1.13}{Dandelion}\DP{0.00}{Gray}\DP{0.75}{Gray}\DP{0.00}{Gray}\DP{0.23}{Purple}\DP{0.00}{Gray}\DP{0.38}{Gray}\DP{0.00}{Gray}\DP{0.60}{Dandelion}\DP{0.00}{Gray}\DP{0.53}{Gray}\DP{0.00}{Gray}\DP{0.30}{Dandelion}\DP{0.00}{Gray}\DP{0.57}{Gray}\\
&CMPL&\DP{1.51}{Gray}\DP{0.00}{Gray}\DP{0.08}{RoyalBlue}\DP{0.00}{Gray}\DP{0.38}{Gray}\DP{0.00}{Gray}\DP{3.32}{RoyalBlue}\DP{0.00}{Gray}\DP{0.23}{Gray}\DP{0.00}{Gray}\DP{0.08}{Dandelion}\DP{0.00}{Gray}\DP{0.08}{RoyalBlue}\DP{0.00}{Gray}\DP{0.98}{Dandelion}\DP{0.00}{Gray}\DP{0.75}{Gray}\DP{0.00}{Gray}\DP{0.23}{Purple}\DP{0.00}{Gray}\DP{0.38}{Gray}\DP{0.00}{Gray}\DP{0.60}{Dandelion}\DP{0.00}{Gray}\DP{0.53}{Gray}\DP{0.00}{Gray}\DP{0.30}{Dandelion}\DP{0.00}{Gray}\DP{0.57}{Gray}\\
&SVFormer&\DP{2.15}{Gray}\DP{0.00}{Gray}\DP{0.15}{Dandelion}\DP{0.00}{Gray}\DP{0.12}{Gray}\DP{0.00}{Gray}\DP{0.08}{Dandelion}\DP{0.00}{Gray}\DP{0.15}{Gray}\DP{0.00}{Gray}\DP{0.13}{RoyalBlue}\DP{0.00}{Gray}\DP{0.03}{Dandelion}\DP{0.00}{Gray}\DP{0.38}{RoyalBlue}\DP{0.00}{Gray}\DP{0.03}{Dandelion}\DP{0.00}{Gray}\DP{1.75}{RoyalBlue}\DP{0.00}{Gray}\DP{0.03}{Dandelion}\DP{0.00}{Gray}\DP{0.15}{RoyalBlue}\DP{0.00}{Gray}\DP{0.05}{Dandelion}\DP{0.00}{Gray}\DP{0.28}{Gray}\DP{0.00}{Gray}\DP{0.93}{Dandelion}\DP{0.00}{Gray}\DP{0.03}{RoyalBlue}\DP{0.00}{Gray}\DP{0.20}{Dandelion}\DP{0.00}{Gray}\DP{0.73}{Gray}\DP{0.00}{Gray}\DP{0.15}{Purple}\DP{0.00}{Gray}\DP{0.40}{Gray}\DP{0.00}{Gray}\DP{0.55}{Dandelion}\DP{0.00}{Gray}\DP{0.03}{RoyalBlue}\DP{0.00}{Gray}\DP{0.07}{Dandelion}\DP{0.00}{Gray}\DP{0.55}{Gray}\DP{0.00}{Gray}\DP{0.23}{Dandelion}\DP{0.00}{Gray}\DP{0.68}{Gray}\\
&DIST&\DP{2.24}{Gray}\DP{0.00}{Gray}\DP{2.97}{RoyalBlue}\DP{0.00}{Gray}\DP{0.28}{Gray}\DP{0.00}{Gray}\DP{1.16}{Dandelion}\DP{0.00}{Gray}\DP{0.70}{Gray}\DP{0.00}{Gray}\DP{0.23}{Purple}\DP{0.00}{Gray}\DP{0.40}{Gray}\DP{0.00}{Gray}\DP{0.60}{Dandelion}\DP{0.00}{Gray}\DP{0.55}{Gray}\DP{0.00}{Gray}\DP{0.23}{Dandelion}\DP{0.00}{Gray}\DP{0.64}{Gray}\\

\hdashline

\multirow{6}{*}{1/16}&Supervised&\DP{2.04}{Gray}\DP{0.00}{Gray}\DP{0.23}{Dandelion}\DP{0.00}{Gray}\DP{0.15}{RoyalBlue}\DP{0.00}{Gray}\DP{0.53}{Dandelion}\DP{1.15}{Gray}\DP{0.21}{RoyalBlue}\DP{0.00}{Gray}\DP{0.08}{Dandelion}\DP{0.00}{Gray}\DP{0.38}{RoyalBlue}\DP{0.00}{Gray}\DP{0.08}{Dandelion}\DP{0.00}{Gray}\DP{0.38}{RoyalBlue}\DP{0.00}{Gray}\DP{0.08}{Dandelion}\DP{0.00}{Gray}\DP{0.23}{Gray}\DP{0.00}{Gray}\DP{0.83}{Dandelion}\DP{0.00}{Gray}\DP{0.15}{RoyalBlue}\DP{0.00}{Gray}\DP{0.15}{Dandelion}\DP{0.00}{Gray}\DP{0.75}{Gray}\DP{0.00}{Gray}\DP{0.15}{Purple}\DP{0.00}{Gray}\DP{0.45}{Gray}\DP{0.00}{Gray}\DP{0.60}{Dandelion}\DP{0.00}{Gray}\DP{0.53}{Gray}\DP{0.00}{Gray}\DP{0.30}{Dandelion}\DP{0.00}{Gray}\DP{0.53}{Gray}\DP{0.00}{Gray}\DP{0.04}{Purple}\\
&FixMatch&\DP{2.57}{Gray}\DP{0.00}{Gray}\DP{0.15}{Dandelion}\DP{0.00}{Gray}\DP{2.49}{RoyalBlue}\DP{0.00}{Gray}\DP{0.30}{Gray}\DP{0.00}{Gray}\DP{1.13}{Dandelion}\DP{0.00}{Gray}\DP{0.75}{Gray}\DP{0.00}{Gray}\DP{0.15}{Purple}\DP{0.00}{Gray}\DP{0.45}{Gray}\DP{0.00}{Gray}\DP{0.60}{Dandelion}\DP{0.00}{Gray}\DP{0.53}{Gray}\DP{0.00}{Gray}\DP{0.30}{Dandelion}\DP{0.00}{Gray}\DP{0.53}{Gray}\DP{0.00}{Gray}\DP{0.04}{Purple}\\
&MeanTeacher&\DP{2.72}{Gray}\DP{0.00}{Gray}\DP{2.49}{RoyalBlue}\DP{0.00}{Gray}\DP{0.38}{Gray}\DP{0.00}{Gray}\DP{0.53}{Dandelion}\DP{0.00}{Gray}\DP{0.08}{Gray}\DP{0.00}{Gray}\DP{0.15}{Dandelion}\DP{0.00}{Gray}\DP{0.15}{Gray}\DP{0.00}{Gray}\DP{0.15}{Dandelion}\DP{0.00}{Gray}\DP{1.43}{Gray}\DP{0.00}{Gray}\DP{0.53}{Dandelion}\DP{0.00}{Gray}\DP{0.60}{Gray}\DP{0.00}{Gray}\DP{0.23}{Dandelion}\DP{0.00}{Gray}\DP{0.57}{Gray}\\
&CMPL&\DP{1.96}{Gray}\DP{0.00}{Gray}\DP{3.24}{RoyalBlue}\DP{0.00}{Gray}\DP{0.30}{Gray}\DP{0.00}{Gray}\DP{0.08}{RoyalBlue}\DP{0.00}{Gray}\DP{0.75}{Dandelion}\DP{0.00}{Gray}\DP{0.08}{RoyalBlue}\DP{0.00}{Gray}\DP{0.30}{Dandelion}\DP{0.00}{Gray}\DP{0.68}{Gray}\DP{0.00}{Gray}\DP{0.15}{Dandelion}\DP{0.00}{Gray}\DP{0.45}{Gray}\DP{0.00}{Gray}\DP{0.60}{Dandelion}\DP{0.00}{Gray}\DP{0.53}{Gray}\DP{0.00}{Gray}\DP{0.30}{Dandelion}\DP{0.00}{Gray}\DP{0.57}{Gray}\\
&SVFormer&\DP{1.95}{Gray}\DP{0.00}{Gray}\DP{0.03}{Dandelion}\DP{0.00}{Gray}\DP{0.05}{Gray}\DP{0.00}{Gray}\DP{0.05}{Dandelion}\DP{0.00}{Gray}\DP{0.05}{Gray}\DP{0.00}{Gray}\DP{0.05}{Dandelion}\DP{0.00}{Gray}\DP{0.50}{Gray}\DP{0.00}{Gray}\DP{0.08}{Dandelion}\DP{0.00}{Gray}\DP{0.13}{Gray}\DP{0.00}{Gray}\DP{0.18}{Dandelion}\DP{0.00}{Gray}\DP{0.08}{RoyalBlue}\DP{0.00}{Gray}\DP{0.10}{Gray}\DP{0.00}{Gray}\DP{1.95}{RoyalBlue}\DP{0.00}{Gray}\DP{0.30}{Gray}\DP{0.00}{Gray}\DP{1.13}{Dandelion}\DP{0.00}{Gray}\DP{0.73}{Gray}\DP{0.00}{Gray}\DP{0.18}{Dandelion}\DP{0.00}{Gray}\DP{0.38}{Gray}\DP{0.00}{Gray}\DP{0.65}{Dandelion}\DP{0.00}{Gray}\DP{0.55}{Gray}\DP{0.00}{Gray}\DP{0.23}{Dandelion}\DP{0.00}{Gray}\DP{0.68}{Gray}\\
&DIST&\DP{2.59}{Gray}\DP{0.00}{Gray}\DP{0.08}{Dandelion}\DP{0.00}{Gray}\DP{2.54}{RoyalBlue}\DP{0.00}{Gray}\DP{0.28}{Gray}\DP{0.00}{Gray}\DP{1.16}{Dandelion}\DP{0.00}{Gray}\DP{0.73}{Gray}\DP{0.00}{Gray}\DP{0.23}{Purple}\DP{0.00}{Gray}\DP{0.38}{Gray}\DP{0.00}{Gray}\DP{0.60}{Dandelion}\DP{0.00}{Gray}\DP{0.55}{Gray}\DP{0.00}{Gray}\DP{0.23}{Dandelion}\DP{0.00}{Gray}\DP{0.64}{Gray}\\

\specialrule{.12em}{.05em}{.05em}
\end{tabular}}
\resizebox{0.5\textwidth}{!}{%
\begin{tabular}{m{1.5cm}m{18cm}}
Colormap:\hspace{1cm} &  Irrigation-Aspiration \DP{0.9}{Dandelion}, \hspace{1cm}
Lens Implantation \DP{0.9}{Purple},\hspace{1cm}
Phako \DP{0.9}{RoyalBlue},\hspace{1cm}
Rest \DP{0.9}{Gray}
\\
\specialrule{.12em}{.05em}{.05em}
\end{tabular}
}
\end{table}

For further evaluations, we tested all the models on a full cataract surgery video, dividing it into three-second clips with a one-second overlap and predicting the label associated with each clip. Table \ref{tab:phase-annotations} presents a subjective assessment of phase prediction performance for different models across two data splits: 1/4 and 1/16. In both splits, the semi-supervised models outperform the supervised model. Remarkably, in the 1/16 split, the DIST model accurately predicts all phases, including the challenging lens implantation phase,  which is shorter and has fewer training examples. While MeanTeacher, CMPL, and SVFormer struggle with this phase, our model identifies it successfully, highlighting its capability to handle complex and underrepresented phases effectively.

\vspace{-1em}

\section{Conclusion}
\label{sec:conclusion}
\vspace{-0.9em}
We introduce DIST (Dual Invariance Self-Training), a novel semi-supervised learning framework for surgical phase recognition. DIST’s core innovation, dual invariance filtering, applies temporal and transformation invariance to select the most reliable pseudo-labels, effectively reducing noise and enhancing pseudo-label quality. Our extensive evaluations on the Cataract-1k and Cholec80 datasets across various network architectures confirm that DIST consistently outperforms both supervised and state-of-the-art SSL models, especially in data-scarce scenarios, establishing it as a robust approach for surgical video analysis. Moreover, DIST is applicable to general video analysis beyond medical domains.

\textbf{Acknowledgements.} This work was funded by the FWF Austrian Science Fund under grant P 32010-N38.

% \vspace{-1.5em}
%\subsubsection*{Acknowledgements} 

%%%%%%%%%%%%%%%%%%%%%%%%%%%%%%%%%%%%%%%%%%%%%%%%%%%%%%
% References
\bibliographystyle{ieeetr}
\bibliography{References}

\end{document}